# Blackbox identity testing for bounded top fanin depth-3 circuits: the field doesn't matter

Nitin Saxena
Hausdorff Center for Mathematics
Bonn, Germany
ns@hcm.uni-bonn.de

C. Seshadhri
Sandia National Laboratories[*]
Livermore, USA
scomand@sandia.gov

**Abstract**

Let $C$ be a depth-3 circuit with $n$ variables, degree $d$ and top fanin $k$ (called $\Sigma\Pi\Sigma(k,d,n)$ circuits) over base field $\mathbb{F}$. It is a major open problem to design a deterministic polynomial time blackbox algorithm that tests if $C$ is identically zero. Klivans & Spielman (STOC 2001) observed that the problem is open even when $k$ is a constant. This case has been subjected to a serious study over the past few years, starting from the work of Dvir & Shpilka (STOC 2005).

We give the first polynomial time blackbox algorithm for this problem. Our algorithm runs in time $\text{poly}(n)d^k$, regardless of the base field. The *only* field for which polynomial time algorithms were previously known is $\mathbb{F} = \mathbb{Q}$ (Kayal & Saraf, FOCS 2009, and Saxena & Seshadhri, FOCS 2010). This is the first blackbox algorithm for depth-3 circuits that does not use the rank based approaches of Karnin & Shpilka (CCC 2008).

We prove an important tool for the study of depth-3 identities. We design a blackbox polynomial time transformation that reduces the number of variables in a $\Sigma\Pi\Sigma(k,d,n)$ circuit to $k$ variables, but preserves the identity structure.

# 1 Introduction

Polynomial identity testing (PIT) is a major open problem in theoretical computer science. The input is an arithmetic circuit that computes a polynomial $p(x_1, x_2, \ldots, x_n)$ over a base field $\mathbb{F}$. We wish to check if $p$ is the zero polynomial, or in other words, is identically zero. We may be provided with an explicit circuit, or may only have blackbox access. In the latter case, we can only evaluate the polynomial $p$ at various domain points. The main goal is to devise a *deterministic* blackbox polynomial time algorithm for PIT. One of the main reasons for interest in this problem is the connection between PIT algorithms and circuit lower bounds (Heintz & Schnorr [HS80], Kabanets & Impagliazzo [KI03] and Agrawal [Agr05, Agr06]). Refer to surveys for a detailed treatment of PIT [Sax09, AS09].

Since the problem of PIT is very hard, restricted versions of it have been studied. One common and natural variant is that of the *bounded depth circuits*. Results of Agrawal & Vinay [AV08] justify this restriction. They essentially show that an efficient blackbox identity test for depth-4 circuits leads to (almost) the complete resolution of PIT and also provides exponential lower bounds. Raz [Raz10] showed that even lower bounds for depth-3 circuits imply super-polynomial lower bounds for general arithmetic formulas. Not surprisingly, the problem of PIT is still wide open for the special case of depth-3 circuits.

A depth-3 circuit $C$ over a field $\mathbb{F}$ is of the form $C(x_1, \ldots, x_n) = \sum_{i=1}^{k} T_i$, where $T_i$ (a *multiplication term*) is a product of at most $d$ linear polynomials with coefficients in $\mathbb{F}$. The size of the circuit $C$ can be expressed in three parameters: the number of variables $n$, the degree $d$, and the *top fanin* $k$. Such a circuit is referred to as a $\Sigma\Pi\Sigma(k, d, n)$ circuit. Even when the top fanin $k$ is constant, blackbox polynomial time algorithms were not known[1].

The study of PIT algorithms for depth-3 circuits was initiated by Dvir & Shpilka [DS05], who gave a quasi-polynomial time non-blackbox algorithm. The first non-trivial *blackbox* algorithm was given by Karnin & Shpilka [KS08]. There have been many recent results in this area by Kayal & Saxena [KS06], Saxena & Seshadhri [SS09, SS10a], and Kayal & Saraf [KS09b]. Our main result is the first polynomial time blackbox tester for bounded top fanin depth-3 circuits over *any* field.

**Theorem 1** *There exists a deterministic blackbox poly($nd^k$) time algorithm for PIT on $\Sigma\Pi\Sigma(k, d, n)$ circuits, regardless of the base field $\mathbb{F}$.*

Table 1 details the time complexities[2] of previous algorithms. For convenience, we do not give the list of all algorithms, but only the important milestones for the case of arbitrary fields. We stress that the time complexities bound the number of bit operations.

Table 1: Depth-3 blackbox PIT algorithms over any field

| Paper | Time complexity |
|---|---|
| [KS08] | $nd^{(2^{k^2} \log^{k-2} d)}$ |
| [SS09, SS10a] | $nd^{k^2 \log d}$ |
| This paper | $nd^k$ |

The only field for which such polynomial time algorithms were previously known was $\mathbb{Q}$. This was a breakthrough result of Kayal & Saraf [KS09b], which was followed by improvements in [SS10a]. These used beautiful incidence geometry theorems for the reals, but analogues of these results are either unknown or false for other fields. Since the best running time of these algorithms is poly($nd^{k^2}$), we get an improved algorithm for this case as well.

As Table 1 shows, *even for the simple case of $k = 3$ and $\mathbb{F} = \mathbb{F}_2$, no deterministic polynomial-time blackbox PIT algorithm was known.* Kayal & Saxena [KS06] gave a non-blackbox algorithm (over all fields), which runs in poly($nd^k$) time [KS06]. Theorem 1 closes the gap between blackbox and non-blackbox algorithms.

---

[1] Until this work.
[2] Technically, the running times are polynomial in the stated times.

Throughout the following discussion, we will think of $k$ as a constant. Hence, when we refer to polynomial time, the dependence on $k$ will be ignored.

## 1.1 Variable reduction

Dvir & Shpilka [DS05] introduced a powerful idea. They defined the notion of the *rank* of a $\Sigma\Pi\Sigma(k,d,n)$ circuit. We will not explain this precisely here but merely say that this is the number of "free variables" in a $\Sigma\Pi\Sigma$ circuit. They proved the remarkable fact that the rank of every identity [3] is small. This led to the reduction of PIT for general $\Sigma\Pi\Sigma(k,d,n)$ circuits to PIT on $\Sigma\Pi\Sigma$ circuits over few variables. They developed a non-blackbox quasi-polynomial time algorithm through their rank bounds. Karnin & Shpilka [KS08] used the idea of rank to devise *blackbox* algorithms for $\Sigma\Pi\Sigma$ circuits. Their algorithms had a running time that depended exponentially in the rank. Hence, constant rank bounds would lead to polynomial time algorithms. Unfortunately, Kayal & Saxena [KS06] gave constructions (extended in [SS09]) showing that for $\Sigma\Pi\Sigma$ identities over finite fields, the rank could be unbounded (as large as $k \log d$). This means that the best running time one could hope for over finite fields via this approach was $d^{k \log d}$. Tighter rank bounds from [SS09, SS10a] gave algorithms that almost match this running time. For the special case when the field $\mathbb{F}$ is $\mathbb{Q}$, Kayal & Saraf [KS09b] proved a constant rank bound, establishing the first polynomial time blackbox algorithm for this case. Refer to [SS10a] for a more detailed treatment of rank bounds.

Until this work, all blackbox algorithms relied solely on the rank approach of Karnin & Shpilka [KS08]. As the examples of [KS06] show, even for the case of $\mathbb{F}_2$, a new idea is required to get polynomial time algorithms. We provide the first blackbox algorithm that circumvents the problem of large rank identities. Interestingly, one of the main ideas has roots in the non-blackbox polynomial time algorithm of Kayal & Saxena [KS06]. This algorithm had a completely different idea and used generalizations of the *Chinese Remainder Theorem*. These algebraic ideas were further developed in a previous work of the authors [SS10a]. Karnin & Shpilka [KS09a] used extractors of Gabizon & Raz [GR05] to construct their blackbox algorithm. We combine these extractor ideas with the algebraic framework to develop a very useful algorithmic tool. Any $\Sigma\Pi\Sigma(k,d,n)$ circuit can be converted into a small family of $\Sigma\Pi\Sigma$ circuits over just $k$ variables. The original circuit is an identity iff all circuits in the family are identities. This transformation requires no knowledge of the circuit and has a running time of poly($kdn$).

**Theorem 2** *Let $\mathbb{F}$ be an arbitrary field such that $|\mathbb{F}| > dnk^2$. There is a deterministic algorithm that takes as input a triple $(k,d,n)$ of natural numbers and in time poly($kdn$), outputs a set of linear maps $\Psi_i : \mathbb{F}[x_1,\ldots,x_n] \to \mathbb{F}[y_1,\ldots,y_k]$ $(1 \leq i \leq poly(kdn))$. A $\Sigma\Pi\Sigma(k,d,n)$ circuit $C$ is identically zero iff $\forall i, \Psi_i(C) = 0$.*

**Remark:** The size restriction made in the theorem is really no loss of generality. In the blackbox model, it is standard to assume that we can query the given circuit on points in an extension field. If $\mathbb{F}$ is small, then we just need to move to a large enough extension field $\mathbb{F}'$. Such an extension can be found by a deterministic poly($\log |\mathbb{F}'|$) time method of Adleman & Lenstra [AL86], or even by a slower brute-force method of finding a suitable irreducible polynomial over $\mathbb{F}$. We do all our computations in this extension field $\mathbb{F}'$. Henceforth, for convenience, we will assume that the field has size $> dnk^2$.

Observe the power of this theorem. Regardless of $n, d$ or $\mathbb{F}$, every $\Sigma\Pi\Sigma(3,d,n)$ non-identity can be converted to an "equivalent" non-identity involving just 3 variables. The time required to generate this transformation is truly polynomial in the *size* of the circuit. We believe that this theorem will be useful in reaching the holy grail of a truly polynomial time PIT algorithm for $\Sigma\Pi\Sigma$ circuits.

---

[3] A small caveat: there are some technical restrictions of simplicity and minimality.

This theorem has a uniform treatment of all fields, and is hence stronger than rank bounds. The circuits $\Psi_i(C)$ only involve $k$ variables. Using some standard PIT techniques, we can construct the following hitting set. This proves Theorem 1.

**Theorem 3** *Given the triple of natural numbers $(k, d, n)$, a hitting set $\mathcal{H} \subseteq \mathbb{F}^n$ for $\Sigma\Pi\Sigma(k, d, n)$ circuits can be constructed in deterministic poly$(nd^k)$ time. In other words, for every non-zero $\Sigma\Pi\Sigma(k, d, n)$ circuit $C$ over $\mathbb{F}$, there is some vector $(\alpha_1, \ldots, \alpha_n) \in \mathcal{H}$ such that $C(\alpha_1, \ldots, \alpha_n) \neq 0$.*

We make a somewhat philosophical remark. By Schwartz-Zippel we know that PIT for a depth-3 circuit $C(x_1, ..., x_n)$ can be done by feeding an $n$-wise independent random distribution. The proof of Theorem 2 shows that for $\Sigma\Pi\Sigma(k, d, n)$ circuits, $k$-wise independent random distribution suffices.

## 1.2 History

The first randomized polynomial time PIT algorithm was given (independently) by Schwartz [Sch80] and Zippel [Zip79]. Algorithms using less randomness were devised by Chen & Kao [CK97], Lewin & Vadhan [LV98], and Agrawal & Biswas [AB99]. For depth-2 circuits, there has been a long line of work studying blackbox PIT algorithms [BOT88, CDGK91, Wer94, KS96, SS96, GKS90, KS01]. Raz & Shpilka studied non-blackbox algorithms for non-commutative formulas [RS05].

Klivans & Spielman [KS01] first observed that deterministic PIT was open even for depth-3 circuits with bounded top fanin. Progress towards this was first made by the quasi-polynomial time algorithm of Dvir & Shpilka [DS05]. The problem was resolved (in the non-blackbox setting) by a polynomial time algorithm given by Kayal & Saxena [KS06], with a running time exponential in the top fanin.

The remaining history of depth-3 PIT has been explained quite a bit in the previous sections. Identity tests are known only for very special depth-4 circuits [AM07, Sax08, SV09, KMSV10]. Agrawal & Vinay [AV08] showed that an efficient blackbox identity test for depth-4 circuits will actually give a quasi-polynomial blackbox test, and exponential lower bounds, for circuits of *all depths* that compute *low degree* polynomials. Thus, understanding depth-3 identities seems to be a natural first step towards the goal of PIT and circuit lower bounds.

At the end, we would just like to indicate how all the depth-3 PIT results are interrelated and how they collectively influenced progress in this problem. The rank notion of Dvir & Shpilka [DS05], the Chinese Remaindering of Kayal & Saxena [KS06], the rank preserving subspaces of Karnin & Shpilka [KS09a], the series of improved rank bounds by the authors and Kayal & Saraf [SS09, KS09b, SS10a]: each paper built of previous results and provided enough food for thought for subsequent papers. This paper also builds on the edifice constructed so far.

## 1.3 Organization

In Section 2, we give some basic definitions and give an intuitive overview of our approach. Section 3 gives some of the tools that were developed in previous works. In Section 4, we give our main analysis and prove Theorems 2 and 3.

## 2 Preliminaries and Intuition

We will denote the set $\{1, \ldots, n\}$ by $[n]$. We fix the base field to be $\mathbb{F}$, so the circuits compute multivariate polynomials in the *polynomial ring* $\mathcal{R} := \mathbb{F}[x_1, \ldots, x_n]$. We use $\mathbb{F}^*$ to denote $\mathbb{F} \setminus \{0\}$. For any subset $S \subseteq [k]$, the *sub-circuit* $C_S$ is $\sum_{s \in S} T_s$. For an $i \in \{0, \ldots, k-1\}$, define $[i]' := [k] \setminus [i]$. Conventionally, $[0] := \emptyset$ and $C_\emptyset := 0$.

A *linear form* is a linear polynomial in $\mathcal{R}$ with zero constant term. We will denote the set of all linear forms by $L(\mathcal{R}) := \{\sum_{i=1}^n a_i x_i \mid a_1, \ldots, a_n \in \mathbb{F}\}$. Clearly, $L(\mathcal{R})$ is a vector (or linear) space over $\mathbb{F}$ and that will be quite useful. Much of what we do shall deal with *multi*-sets of linear forms (also product of linear forms) and equivalence classes inside them. A *list* of linear forms is a multi-set of forms with an arbitrary order associated with them.

**Definition 4** *We collect some important definitions from [SS09]:*

[**Multiplication term and operators** $L(\cdot)$ & $M(\cdot)$] *A* multiplication term *$f$ is an expression in $\mathcal{R}$ given as (the product may have repeated $\ell$'s), $f := c \cdot \prod_{\ell \in S} \ell$, where $c \in \mathbb{F}^*$ and $S$ is a list of nonzero linear forms. The* list of linear forms in $f$, $L(f)$, *is just the list $S$ of forms occurring in the product above. For a list $S$ of linear forms we define the* multiplication term of $S$, $M(S)$, *as $\prod_{\ell \in S} \ell$ or 1 if $S = \phi$.*

[**ΣΠΣ circuits**] *An $\Sigma\Pi\Sigma(k, d)$ circuit $C$ is a sum of $k$ multiplication terms of degree $d$, $C = \sum_{i=1}^{k} T_i$. The list of linear forms occurring in $C$ is $L(C) := \bigcup_{i \in [k]} L(T_i)$. Note that $L(C)$ is a list of size exactly $kd$. (For the purposes of this paper $T_i$'s are given in circuit representation and thus the list $L(T_i)$ is unambiguously defined from $C$.)*

[**Span** $sp(\cdot)$ **and Rank** $rk(\cdot)$] *For any $S \subseteq L(\mathcal{R})$ we let $sp(S) \subseteq L(\mathcal{R})$ be the* linear span *of the linear forms in $S$ over the field $\mathbb{F}$. (Conventionally, $sp(\emptyset) = \{0\}$.) We use $rk(S)$ to denote the rank of $S$, considered as vectors in $\mathbb{F}^n$.*

## 2.1 Intuition and main ideas

We give a high-level description of the main ideas used in this paper. Some notions are deliberately left vague, and others may even be formally incorrect. Nonetheless, this sketch is "morally" correct and, at some level, shows how the authors arrived at their conclusions.

How do we convert a high variate $\Sigma\Pi\Sigma(k, d, n)$ circuit $C$ into a low variate one and still preserve the structure of $C$? We wish to do this by a linear transformation $\Psi : \mathbb{F}[x_1, \ldots, x_n] \to \mathbb{F}[y_1, \ldots, y_\ell]$, where $\ell$ is comparable to $k$. When the rank of the linear forms in $C$ is itself comparable to $k$, this can be done quite directly. We will get a circuit $\Psi(C)$ that is essentially *isomorphic* to $C$. For identities of large rank, such a transformation seems impossible. Any linear transformation will necessarily destroy some of the structure of $C$. This is because forms that were independent in $C$ are now dependent in $\Psi(C)$. But maybe we are trying too hard to preserve the dependencies in $C$? After all, we want a transformation $\Psi$ that sends identities to identities. (And, of course, non-identities to non-identities. Surely, satisfying only the former condition is not too hard.) We are not particularly bothered about how well $\Psi(C)$ preserves the exact structure of $C$.

This is where the Chinese Remaindering techniques of [KS06] and the ideal framework of [SS10a] enters the picture. For any non-identity $C$, it was shown that there exists an ideal $I$ generated by products of forms in $L(C)$ that "certifies" that $C$ is non-zero. Essentially, the polynomial $C$ is not in ideal $I$ and hence, must be non-zero. In reality, it is much more complicated than that, but for the sake of explanation, it captures the main idea. The forms involved in generating $I$ have rank at most $k$. This gives a low-dimensional certificate of the non-zeroness of $C$.

We argue that if $\Psi$ can selectively preserve the forms generating $I$, then $\Psi(I)$ remains a certificate for $\Psi(C)$. In other words, $\Psi(C)$ will not be in $\Psi(I)$. All that is needed is, to find such a $\Psi$ that is independent of $C$ (since we are interested in blackbox algorithms). Enter [KS09a]. They develop the notion of *rank-preserving subspaces*. These can be viewed as linear transformations from a large-dimensional vector space $\mathcal{S}_1$ to a smaller dimensional one $\mathcal{S}_2$. These preserve the structure (in terms of linear independence) of specific low dimensional subspaces of $\mathcal{S}_1$. Furthermore, they can be constructed in a blackbox manner using extractors of [GR05]. To make this work, we will actually need a *set* of transformations, and one $\Psi$ will not suffice.

The circuit $\Psi(C)$ is of the form $\Sigma\Pi\Sigma(k, d, k)$. This has only a constant number of variables, and the Schwartz-Zippel lemma [Sch80, Zip79] gives a simple blackbox algorithm for such circuits. This is combined with the transformation $\Psi$ to construct the final hitting set.

# 3 Necessary tools

In this section, we list out the basic tools that we need. In the first part, we explain the low-rank ideal certificates for the non-zeroness of $C$. This requires some technical definitions, before we can state the exact theorem. In the second part, we give the key lemma of the Vandermonde matrix transformation used in [KS09a]. Once these tools are set in place, we will explain how the variable reduction works.

## 3.1 Low rank certificates

This framework and the following theorems were developed in [SS10a]. The definitions are extremely technical, and may appear to be somewhat unmotivated. These are needed to precisely formalize the notion of the low-rank certificate for non-zeroness. We reproduce many of the definitions and details for convenience. For more details, the interested reader should see the full version of [SS10a] ([SS10b]).

**Definition 5 (Ideal)** *An* ideal $I$ *of* $\mathcal{R}$ *with generators* $f_i, i \in [m]$, *is the set* $\{\sum_{i \in [m]} q_i f_i | q_i\text{'s} \in \mathcal{R}\}$ *and is denoted by the notation* $\langle f_1, \ldots, f_m \rangle$. *For an* $f \in \mathcal{R}$, *the three notations* $f \equiv 0 (mod\ I)$, $f \equiv 0 (mod\ f_1, \ldots, f_m)$ *and* $f \in I$, *mean the same.*

**(Radical-span)** *Let* $S := \{f_1, \ldots, f_m\}$ *be multiplication terms generating an ideal* $I$. *We associate a linear space to* $S$ *called the* radical span, $radsp(S) := sp(L(f_1) \cup \ldots \cup L(f_m))$.

*When the set of generators* $S$ *are clear from the context we will also use the notation* $radsp(I)$. *Similarly,* $radsp(I, f)$ *would be a shorthand for* $radsp(S \cup \{f\})$.

**(Nodes)** *Let* $f$ *be a multiplication term and let* $I$ *be an ideal generated by some multiplication terms. As the relation "similarity mod* $radsp(I)$*" is an equivalence relation on* $L(\mathcal{R})$, *it partitions the list* $L(f)$ *into equivalence classes.*

[$\mathbf{rep}_I(f)$] *For each such class, pick a representative* $\ell_i$ *and define their collection* $rep_I(f) := \{\ell_1, \ldots, \ell_r\}$. *(Note that form* $0$ *can also appear in this set, it represents the class* $L(f) \cap radsp(I)$.*)*

[$\mathbf{nod}_I(f)$] *For each* $\ell_i \in rep_I(f)$, *we multiply the forms in* $f$ *that are similar to* $\ell_i$ *mod* $radsp(I)$. *We define nodes of* $f$ *mod* $I$ *as the set of polynomials* $nod_I(f) := \{M(L(f) \cap (\mathbb{F}^*\ell + radsp(I))) \mid \ell \in rep_I(f)\}$. *(Remark: When* $I = \{0\}$, *nodes of* $f$ *are just the coprime powers-of-forms dividing* $f$.*)*

**(Paths)** *Let* $I$ *be an ideal generated by some multiplication terms. Let* $C = \sum_{i \in [k]} T_i$ *be a* $\Sigma\Pi\Sigma(k, d)$ *circuit. Let* $v_i$ *be a sub-term of* $T_i$ *(i.e.* $L(v_i) \subseteq L(T_i)$*), for all* $i \in [k]$. *We call the tuple* $(I, v_1, \ldots, v_k)$ *a path of* $C$ *mod* $I$ *if, for all* $i \in [k]$, $v_i \in nod_{\langle I, v_1, \ldots, v_{i-1} \rangle}(T_i)$. *It is of length* $k$. *(Remark: We have defined path* $\overline{p}$ *as a tuple but, for convenience, we will sometimes treat it as a set of multiplication terms, eg. when operated upon by* $sp(\cdot)$, $\langle \cdot \rangle$, $radsp(\cdot)$, *etc.)*

*Conventionally, when* $k = 0$ *the circuit* $C$ *has just "one" gate:* $0$. *In that case, the only path* $C$ *mod* $I$ *has is* $(I)$, *which is of length* $0$.

Observe that the product of polynomials in $nod_I(f)$ just gives $f$ (upto a constant multiple). Also, modulo $radsp(I)$, each node is just a form-power $\ell^m$. In other words, modulo $radsp(I)$, a node is a rank-one term. Figure 1 should clarify the definition. The oval bubbles represent the list of forms in a term, and the rectangles enclose forms in a node. The arrows show a path. Starting with $I$ as the zero ideal, $v_1 := x_1^2$, $v_2 := x_2(x_2 + 2x_1)$, and $v_3 := (x_4 + x_2)(x_4 + 4x_2 - x_1)(x_4 + x_2 + x_1)(x_4 + x_2 - 2x_1)$ form a path. Initially the path is just the zero ideal, so $x_1^2$ is a node. Note how $v_2$ is a power of $x_2$ modulo $radsp(v_1)$ and $v_3$ is a power of $x_4$ modulo $radsp\langle v_1, v_2 \rangle$.

Theorem 25 of [SS10b] claims that in a non-zero depth-3 circuit $C$, there always exists a path "certifying" the non-zeroness of $C$. Essentially, there is a path $\overline{p}$ such that modulo $\overline{p}$, the circuit $C$ reduces to a single non-zero multiplication term. The theorem is stated for general settings, but we will always assume that $I = \langle 0 \rangle$. Note that the rank of $\mathrm{radsp}(\overline{p})$ is at most $(k-1)$, since each node can increase the rank by at most 1 and since one term is to be left uncancelled.

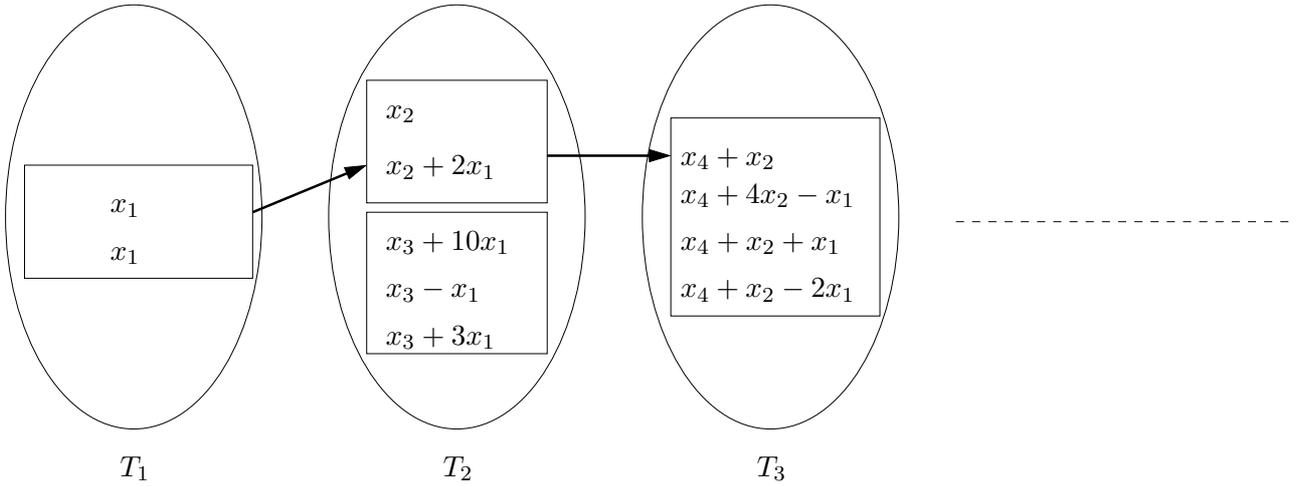

Figure 1: Nodes and paths in $C = T_1 + T_2 + T_3 + \ldots$

**Theorem 6 (Certificate for a Non-identity)** *Let $I$ be an ideal generated by some multiplication terms. Let $C = \sum_{i \in [k]} T_i$ be a $\Sigma\Pi\Sigma(k,d)$ circuit that is nonzero modulo $I$. Then $\exists i \in \{0, \ldots, k-1\}$ such that $C_{[i]} \bmod I$ has a path $\overline{p}$ satisfying: $C_{[i]'} \equiv \alpha \cdot T_{i+1} \not\equiv 0 \pmod{\overline{p}}$ for some $\alpha \in \mathbb{F}^*$.*

## 3.2 The Vandermonde linear transformation

Linear transformations based on the Vandermonde matrix are used widely, for eg. in [GR05] to construct *linear seeded extractors for affine sources*. This was adapted in [KS08] to design depth-3 blackbox PIT algorithms. We will need ideas from Lemma 6.1 from [GR05] to construct linear transformations that reduce the dimension of a space but still preserve the structure of low rank subspaces.

We will design a linear transformation $\Psi_\beta : \mathbb{F}^n \to \mathbb{F}^k$ for any $\beta \in \mathbb{F}$ as follows. Let $V_{n,k,\beta}$ denote the $n \times k$ Vandermonde matrix. This is defined as $(V_{n,k,\beta})_{i,j} := \beta^{ij}$. We have

$$(b_1 \ldots b_k) = (a_1 \ldots a_n) \cdot V_{n,k,\beta} \qquad (b_r = \sum_{i=1}^n a_i \beta^{ri}) \tag{1}$$

So far, $\Psi_\beta$ has been seen as a linear transformation. But we wish to eventually understand its action on ideals. For that reason, it is necessary to view $\Psi_\beta$ as a *linear homomorphism* from $\mathcal{R} = \mathbb{F}[x_1, \ldots, x_n]$ to $\mathcal{R}' := \mathbb{F}[y_1, \ldots, y_k]$. This means that $\Psi_\beta$ maps $\mathcal{R}$ to $\mathcal{R}'$ and preserves the ring operations of addition and multiplication. We can equivalently define $\Psi_\beta$ as

$$\forall i \in [n], \quad \Psi_\beta : x_i \mapsto \sum_{j=1}^k \beta^{ij} y_j \tag{2}$$

We define $\Psi_\beta(\alpha) = \alpha$ for all $\alpha \in \mathbb{F}$. This (naturally) defines the action of $\Psi_\beta$, on *all* the elements of $\mathcal{R}$, that preserves the ring operations of $\mathcal{R}$. We now state a key property of $\Psi$ (Lemma 6.1 of [GR05]). For completeness, a proof is provided in Appendix A.

**Lemma 7 ($\Psi_\beta$ preserves $k$-rank)** *Let $\Psi_\beta : \mathcal{R} \to \mathcal{R}'$ be the linear homomorphism defined by Equation (2). Let $S \subseteq L(\mathcal{R})$ be a subset of linear forms with $rk(S) \leq k$. For all but $nk^2$ values of $\beta$, $rk(\Psi_\beta(S)) = rk(S)$.*

*The homomorphism $\Psi_\beta$ depends solely on the triple of natural numbers $(k, d, n)$ and is computable in poly$(kdn)$ time.*

**Remark:** At first sight $\Psi_\beta$ might seem independent of $d$ but recall that we assume $|\mathbb{F}| > dnk^2$.

# 4 Analysis

With all the basic definitions in place, we are now ready to convert any $\Sigma\Pi\Sigma(k,d,n)$ non-identity $C$ into a $\Sigma\Pi\Sigma(k,d,k)$ non-identity. This will be done through a two-step process. We will use the properties of $\Psi_\beta$ to prove a major generalization of Lemma 7. Not only does $\Psi_\beta$ preserves small subspaces, but it also maintains the structure of ideals having low radical-span. This leads to the second step. We deduce that $\Psi_\beta$ must preserve all paths of $C$, since paths have a low radical span. Since $C$ has a certifying path $p$, we show that $\Psi_\beta(p)$ must also be a certificate for $\Psi_\beta(C)$.

## 4.1 The moral nature of $\Psi_\beta$: it maintains ideals

We study the action of $\Psi_\beta$ on ideals. Our main lemma is the following.

**Lemma 8 ($\Psi_\beta$ preserves ideals)** *Let $f_1, \ldots, f_m, f$ be multiplication terms in $\mathcal{R}$. Define the ideal $I := \langle f_1, \ldots, f_m \rangle$. Let the span $sp(L(f) \cup radsp(I))$, over $\mathbb{F}$, be of rank at most $k$. Then, for all but $nk^2$ values of $\beta$: $f \in I$ iff $\Psi_\beta(f) \in \langle \Psi_\beta(f_1), \ldots, \Psi_\beta(f_m) \rangle$.*

To prove this, we need to show that the map $\Psi_\beta$ is an isomorphism on small enough subrings of $\mathcal{R}$. This is a fairly direct consequence of Lemma 7. For $\ell_1, \ldots, \ell_k \in L(\mathcal{R})$, $\mathbb{F}[\ell_1, \ldots, \ell_k] \subset \mathcal{R}$ denotes the set of all polynomials $g(\ell_1, \ldots, \ell_k)$, where $g \in \mathbb{F}[y_1, \ldots, y_k]$. This is the *subalgebra* of $\mathcal{R}$ generated by $\{\ell_1, \ldots, \ell_k\}$.

**Lemma 9** *Let $\ell_1, \ldots, \ell_k \in L(\mathcal{R})$ be $k$ linearly independent forms. Then, for all but $nk^2$ values of $\beta$, $\Psi_\beta$ induces an isomorphism between $\mathbb{F}[\ell_1, \ldots, \ell_k]$ and $\mathcal{R}'$.*

**Proof.** Let $\mathcal{B}$ denote $\mathbb{F}[\ell_1, \ldots, \ell_k]$ and set $S := \{\ell_1, \ldots, \ell_k\} \subset L(\mathcal{R})$. It will be convenient to define $\Phi_\beta$ to be the map induced by $\Psi_\beta$ on $\mathcal{B}$. We will show that $\Phi_\beta : \mathcal{B} \to \mathcal{R}'$ is an isomorphism. Since $S$ is of rank $k$, $\mathcal{B}$ is isomorphic to $\mathcal{R}'$ (under an invertible linear transformation of variables). To show that the homomorphism $\Phi_\beta$ is an isomorphism, it suffices to prove that $\Phi_\beta$ is onto. We need to show that for any $g(y_1, \ldots, y_k) \in \mathcal{R}'$, there exists $p \in \mathcal{B}$ such that $\Phi_\beta(p) = g$.

For the set $S$, choose a value of $\beta$ other than the $nk^2$ values given by Lemma 7. We have $\text{rk}(\Phi_\beta(S)) = \text{rk}(S) = k$. Therefore, for each $y_i$, there exist constants $\alpha_j \in \mathbb{F}$ such that $y_i = \sum_j \alpha_j \Phi_\beta(\ell_j)$. By the linearity of $\Phi_\beta$, $y_i = \Phi_\beta(\sum_j \alpha_j \ell_j)$. Hence, for each $y_i$, there is some linear form $t_i \in L(\mathcal{B})$, such that $\Phi_\beta(t_i) = y_i$. The polynomial $p := g(t_1, \ldots, t_k)$ is certainly in $\mathcal{B}$. Since $\Phi_\beta$ is a homomorphism, $\Phi_\beta(p) = g$. ∎

We can now complete the proof of Lemma 8.
**Proof.** If $f \in I$, then $f = \sum_{i \in [m]} g_i f_i$, where $g_i \in \mathcal{R}$. Since $\Psi_\beta$ is a homomorphism, $\Psi_\beta(f) = \sum_{i \in [m]} \Psi_\beta(g_i) \Psi_\beta(f_i)$. So $\Psi_\beta(f) \in \langle \Psi_\beta(f_1), \ldots, \Psi_\beta(f_m) \rangle$.

Let the span $sp(L(f) \cup radsp(I))$ over $\mathbb{F}$ be generated by linear forms $\ell_1, \ldots, \ell_r \in L(\mathcal{R})$. Since the rank of $sp(L(f) \cup radsp(I))$ is at most $k$, $r \leq k$. Choose arbitrary forms $\ell_{r+1}, \ldots, \ell_k$ such that $\ell_1, \ldots, \ell_k$ are linearly independent. Define subring (of $\mathcal{R}$) $\mathcal{B} := \mathbb{F}[\ell_1, \ldots, \ell_k]$. Applying Lemma 9, we get that for all but $nk^2$ values of $\beta$, $\Psi_\beta$ induces an isomorphism $\Phi_\beta : \mathcal{B} \to \mathcal{R}'$. Choose any such $\beta$ and fix the unique elements $t_1, \ldots, t_k \in \mathcal{B}$ such that $\Phi_\beta(t_i) = y_i$ for all $i \in [k]$.

Suppose $\Psi_\beta(f) \in \langle \Psi_\beta(f_1), \ldots, \Psi_\beta(f_m) \rangle$. Then there exists $g_1, \ldots, g_m \in \mathcal{R}'$ such that,

$$\Psi_\beta(f) = \sum_{i=1}^m g_i \cdot \Psi_\beta(f_i) \qquad (3)$$

Each $g_i$ is a polynomial in $(y_1, \ldots, y_k)$ over $\mathbb{F}$. So we can define the polynomial,

$$h := f - \sum_{i=1}^m g_i(t_1, \ldots, t_k) \cdot f_i.$$

Note that $f$ and $f_i$'s are multiplication terms generated by forms in the $\text{sp}(\ell_1, \ldots, \ell_k)$. Hence all of these are in $\mathcal{B}$. The polynomials $g_i(t_1, \ldots, t_k)$ are also in $\mathcal{B}$, so $h \in \mathcal{B}$. Since $\Phi_\beta$ is a homomorphism,

$$\Phi_\beta(h) = \Phi_\beta(f) - \sum_{i=1}^{m} g_i(\Phi_\beta(t_1), \ldots, \Phi_\beta(t_k)) \cdot \Phi_\beta(f_i) = \Psi_\beta(f) - \sum_{i=1}^{m} g_i(y_1, \ldots, y_k) \cdot \Psi_\beta(f_i) = 0$$

But $\Phi_\beta$ is an isomorphism, so $h = 0$. This implies $f = \sum_{i=1}^{m} g_i(t_1, \ldots, t_k) \cdot f_i$. Note that the evaluations of the $g_i$'s are in $\mathcal{R}$. Thus, $f \in \langle f_1, \ldots, f_m \rangle$. ∎

## 4.2 Variable reduction

We come to the main part where we merge the path certificates with the properties of $\Psi_\beta$ to prove the variable reduction. We will need a technical cancellation lemma from [SS10b], proven in Appendix B.

**Lemma 10** *Let $f_1, \ldots, f_m$ be multiplication terms generating an ideal $I$, let $\ell \in L(\mathcal{R})$ and $g \in \mathcal{R}$. If $\ell \notin \text{radsp}(I)$ then: $\ell g \in I$ iff $g \in I$.*

We now state our main variable reduction lemma. This, combined with the polynomial time constructions of the $\Psi_\beta$'s, completes the proof of Theorem 2.

**Lemma 11** *Let $C$ be a $\Sigma\Pi\Sigma(k, d, n)$ circuit and $U \subseteq \mathbb{F}$ such that $|U| = dnk^2 + 1$. Then $C = 0$ iff $\forall \beta \in U, \Psi_\beta(C) = 0$.*

**Proof.** Since $\Psi_\beta$ is a homomorphism, $C = 0$ implies $\forall \beta, \Psi_\beta(C) = 0$.

Suppose $C \neq 0$, but $\forall \beta \in U, \Psi_\beta(C) = 0$. Applying Theorem 6 on $C$ (with $I := \langle 0 \rangle$) yields a *certifying path* $\overline{p}$. Thus, $\exists i \in \{0, \ldots, k-1\}$ such that $C_{[i]} \mod I$ has a path $\overline{p}$ satisfying,

$$C_{[i]'} \equiv \alpha \cdot T_{i+1} \not\equiv 0 \pmod{\overline{p}}, \text{ for some } \alpha \in \mathbb{F}^*. \tag{4}$$

Note that $\overline{p}$ is basically a sequence of multiplication terms such that $\text{rk}(\text{radsp}(\overline{p})) < k$. Let $g := M(L(T_{i+1}) \cap \text{radsp}(\overline{p}))$. This is just the product of all forms in $T_{i+1}$ that are in $\text{radsp}(\overline{p})$. Note that $T_{i+1}/g$ is a product of forms *not* in $\text{radsp}(\overline{p})$. By repeated applications of Lemma 10, since $T_{i+1} \notin \langle \overline{p} \rangle$, $g \notin \langle \overline{p} \rangle$. The rank of $\text{sp}(L(g) \cup \text{radsp}(\overline{p}))$ is less than $k$. Indeed, for any linear form $\ell \in L(T_{i+1})$, the rank of $\{\ell\} \cup \text{radsp}(\overline{p})$ is at most $k$.

We will now collect a set $B$ of "bad" $\beta$ values. By Lemma 7, for each $\ell \in L(T_{i+1})$, there are at most $nk^2$ values of $\beta$ such that $\Psi_\beta$ does not preserve $\text{rk}(\{\ell\} \cup \text{radsp}(\overline{p}))$. Add all of these values to $B$. The total number of all these bad $\beta$ values is at most $dnk^2$. Therefore, there exists a good $\beta$ in $U$.

For any $\beta \in U \setminus B$, we know that $\Psi_\beta$ preserves $\text{rk}(L(g) \cup \text{radsp}(\overline{p})) = \text{rk}(\text{radsp}(\overline{p}))$. Thus, by the proof of Lemma 8, $\Psi_\beta$ preserves $g \notin \langle \overline{p} \rangle$. In other words, $\Psi_\beta(g) \notin \langle \Psi_\beta(\overline{p}) \rangle$. We get a contradiction with the following claim.

**Claim 12** *Choose $\beta \in U \setminus B$. If $\Psi_\beta(C) = 0$, then $\Psi_\beta(g) \in \langle \Psi_\beta(\overline{p}) \rangle$.*

**Proof.** Observe that $C_{[i]} \equiv 0 \pmod{\overline{p}}$, implying $\Psi_\beta(C_{[i]}) \equiv 0 \pmod{\Psi_\beta(\overline{p})}$. We get $0 = \Psi_\beta(C) = \Psi_\beta(C_{[i]}) + \Psi_\beta(C_{[i]'})$. Going modulo $\overline{p}$ and applying Equation (4), $\Psi_\beta(T_{i+1}) \equiv 0 \pmod{\Psi_\beta(\overline{p})}$. In terms of ideals, $\Psi_\beta(T_{i+1}) \in \langle \Psi_\beta(\overline{p}) \rangle$. Consider any form $\ell \in L(T_{i+1})$ such that $\ell \notin \text{radsp}(\overline{p})$. We will show that $\Psi_\beta(T_{i+1})/\Psi_\beta(\ell) \in \langle \Psi_\beta(\overline{p}) \rangle$.

We have $\text{rk}(\{\ell\} \cup \text{radsp}(\overline{p})) = \text{rk}(\text{radsp}(\overline{p})) + 1$, by the choice of $\ell$. Since $\beta \notin B$, $\text{rk}(\{\Psi_\beta(\ell)\} \cup \text{radsp}(\Psi_\beta(\overline{p}))) = \text{rk}(\text{radsp}(\Psi_\beta(\overline{p}))) + 1$. This implies $\Psi_\beta(\ell) \notin \text{radsp}(\Psi_\beta(\overline{p}))$. Since $\Psi_\beta(T_{i+1}) \in \langle \Psi_\beta(\overline{p}) \rangle$, Lemma 10 tells us that $\Psi_\beta(T_{i+1})/\Psi_\beta(\ell) \in \langle \Psi_\beta(\overline{p}) \rangle$. We can iteratively repeat this process for all such forms $\ell$. We will end up with $\Psi_\beta(g) \in \langle \Psi_\beta(\overline{p}) \rangle$. ∎

### 4.3 The final hitting set

Let $(k, d, n)$ be the triple of natural numbers given in the input. We design a simple hitting set $\mathcal{H}$. We will generate a set of vectors $\overline{\delta} \in \mathbb{F}^n$ that make $\mathcal{H}$.

- Let $S \subseteq \mathbb{F}$ be an arbitrary set of size $dnk^2 + 1$.
- Let $T \subseteq \mathbb{F}$ be an arbitrary set of size $d + 1$.
- For each $\beta \in S$ and each vector $(\gamma_1, \ldots, \gamma_k) \in T^k$, define the vector $\overline{\delta}$ in $\mathbb{F}^n$ as follows:

$$\delta_i := \sum_{j \in [k]} \beta^{ij} \gamma_j.$$

We will use the classical Schwartz-Zippel theorem.

**Theorem 13** *(Schwartz-Zippel) Let $f(y_1, \ldots, y_k)$ be a polynomial of degree $d$. Let $T$ be a finite subset of $\mathbb{F}$. The probability that $f$ is zero on a random point in $T^k$ is at most $d/|T|$.*

*Thus, for $|T| > d$, $T^k$ is a hitting set for all $k$-variate polynomials of degree $d$.*

We are now all set to finish the proof of Theorem 3.

**Theorem 14** *The set $\mathcal{H}$ is a hitting set for $\Sigma\Pi\Sigma(k, d, n)$ circuits. It can be generated in $poly(nd^k)$ time.*

**Proof.** The latter statement is quite clear, given the construction of $\mathcal{H}$. Consider a non-zero $\Sigma\Pi\Sigma(k, d, n)$ circuit $C$. We need to show the existence of some $\overline{\delta} \in \mathcal{H}$ such that $C(\overline{\delta}) \neq 0$. By Lemma 11, there exists a $\beta \in S$ such that $\Psi_\beta(C) \neq 0$. Since $\Psi_\beta(C)$ is a $\Sigma\Pi\Sigma(k, d, k)$ circuit, Theorem 13 tells us that there is some $\overline{\gamma} \in T^k$ such that $\Psi_\beta(C)(\overline{\gamma}) \neq 0$. Consider the $\overline{\delta}$ corresponding to this $\beta$ and $\overline{\gamma}$. By construction of $\overline{\delta}$ and the definition of $\Psi_\beta$ (Equation 2), $C(\overline{\delta}) = \Psi_\beta(C)(\overline{\gamma}) \neq 0$. ∎

## 5 Conclusion

We show that $\Sigma\Pi\Sigma(k, d, n)$ identity is only as complicated as a $\Sigma\Pi\Sigma(k, d, k)$ identity. We prove this fact by observing that there is a "low rank" homomorphism that preserves the ideal structure in depth-3 circuits. Since this low rank homomorphism is easily computable, we get a poly($nd^k$) time blackbox test. Can we identify properties of $k$-variate fanin $k$ identities to develop faster PIT algorithms? Currently, no PIT algorithm is able to beat the exponential dependence on $k$.

This work also raises a question for depth-4 circuits: are there analogous low rank homomorphisms for $\Sigma\Pi\Sigma\Pi(k)$ circuits? Such results would open the door for interesting PIT algorithms for higher depth circuits.

Can this approach be used beyond PIT? In particular, there are results known about learning $\Sigma\Pi\Sigma(k)$ circuits where PIT methods have turned out to be useful [KS09a]. The variable reduction techniques might have some utility for these problems.

## Acknowledgements

We are grateful to Hausdorff Center for Mathematics, Bonn for its kind support, especially in hosting the second author when part of the work was done. The first author thanks Malte Beecken and Johannes Mittmann for several interesting discussions.

# A  A Vandermonde-inspired Linear Transformation

**Lemma 7.** Let $\Psi_\beta : \mathcal{R} \to \mathcal{R}'$ be the linear homomorphism defined by Equation (2). Let $S \subseteq L(\mathcal{R})$ be a subset of linear forms with $\mathrm{rk}(S) \leq k$. For all but $nk^2$ values of $\beta$, $\mathrm{rk}(\Psi_\beta(S)) = \mathrm{rk}(S)$.

The homomorphism $\Psi_\beta$ depends solely on the triple of natural numbers $(k, d, n)$ and is computable in $\mathrm{poly}(kdn)$ time.

**Proof.** We can assume wlog that $S$ is of rank $k$, and has $k$ linearly independent forms $\ell_1, \ldots, \ell_k \in L(\mathcal{R})$. Say $\Psi_\beta$ maps $\ell_i = \sum_{j \in [n]} a_{i,j} x_j$ to $\sum_{j \in [k]} b_{i,j} y_j$, for all $i \in [k]$. Define matrices $B := ((b_{i,j}))_{i \in [k], j \in [k]}$ and $A := ((a_{i,j}))_{i \in [k], j \in [n]}$. By Equation (1) we deduce $B = A \cdot V_{n,k}$. We will show that $B$ is invertible.

Since the rows of $A$ are linearly independent over $\mathbb{F}$, we can apply *partial* Gaussian elimination (i.e. row operations) on $A$. This has the effect of *left*-multiplying $A$ by an invertible matrix $E \in \mathbb{F}^{k \times k}$ ensuring: there are column indices $j_1 > \ldots > j_k \in [n]$ such that $j_i$ is the *maximal* index with $(EA)_{i,j_i} \neq 0$, for all $i \in [k]$. Now we consider the matrix $B' := (EA) \cdot V_{n,k}$,

$$\det(B') = \sum_{\sigma \in S_k} \mathrm{sgn}(\sigma) \cdot P_\sigma(\beta), \quad \text{where } P_\sigma(\beta) := \prod_{i \in [k]} B'_{i, \sigma(i)}.$$

Note that we view $B'_{i,\sigma(i)}$ as a polynomial in $\mathbb{F}[\beta]$ which, by the assumption on $EA$, is of degree $j_i \cdot \sigma(i)$. Thus,

$$\deg(P_\sigma(\beta)) = \sum_{i \in [k]} j_i \cdot \sigma(i).$$

Since $j_1 > \ldots > j_k$, it can be easily shown that the expression above achieves its maxima (over $\sigma \in S_k$) only if $\sigma(1) > \ldots > \sigma(k)$. But this uniquely specifies $\sigma$, hence there is a unique $P_\sigma(\beta)$ of the largest degree ($\leq nk^2$). Thus, $\det(B')$ is a nonzero polynomial in $\mathbb{F}[\beta]$ of degree at most $nk^2$. This means that $B'$ is invertible, hence $B = E^{-1} B'$ is invertible, for all but at most $nk^2$ values of $\beta$. ∎

# B  A Cancellation Lemma

An $f \in \mathcal{R}$ is called a *zerodivisor* of an ideal $I$ (or mod $I$) if $f \notin I$ and there exists a $g \in \mathcal{R} \setminus I$ such that $fg \in I$.

Let $u, v \in \mathcal{R}$. It is easy to see that if $u$ is nonzero mod $I$ and is a *non*-zerodivisor mod $I$ then: $uv \in I$ iff $v \in I$. This can be seen as some sort of a "cancellation rule" for non-zerodivisors. We show such a cancellation rule in the case of ideals arising in $\Sigma\Pi\Sigma$ circuits.

**Lemma 10.** Let $f_1, \ldots, f_m$ be multiplication terms generating an ideal $I$, let $\ell \in L(\mathcal{R})$ and $g \in \mathcal{R}$. If $\ell \notin \mathrm{radsp}(I)$ then: $\ell g \in I$ iff $g \in I$.

**Proof.** Assume $\ell \notin \mathrm{radsp}(I)$. If $I = \{0\}$ then the lemma is of course true. So let us assume that $I \neq \{0\}$ and $\mathrm{rk}(\mathrm{radsp}(I)) =: r \in [n-1]$. As $\ell \notin \mathrm{radsp}(I)$ there exists an invertible linear transformation $\tau : L(\mathcal{R}) \to L(\mathcal{R})$ that maps each form of $\mathrm{radsp}(I)$ to $\mathrm{sp}(x_1, \ldots, x_r)$ and maps $\ell$ to $x_n$. Now suppose that $\ell g \in I$. This means that there are $q_1, \ldots, q_m \in \mathcal{R}$ such that $\ell g = \sum_{i=1}^m q_i f_i$. Apply $\tau$ on this to get:

$$x_n g' = \sum_{i=1}^m q_i' \tau(f_i). \tag{5}$$

We know that $\tau(f_i)$'s are free of $x_n$. Express $g', q_i'$-s as polynomials wrt $x_n$, say

$$g' = \sum_{j \geq 0} a_j x_n^j, \text{ where } a_j \in \mathbb{F}[x_1, \ldots, x_{n-1}] \tag{6}$$

$$q_i' = \sum_{j \geq 0} b_{i,j} x_n^j, \text{ where } b_{i,j} \in \mathbb{F}[x_1, \ldots, x_{n-1}] \tag{7}$$

Now for some $d \geq 1$ compare the coefficients of $x_n^d$ on both sides of Equation (5). We get $a_{d-1} = \sum_{i=1}^m b_{i,d} \tau(f_i)$, thus $a_{d-1}$ and $a_{d-1} x_n^{d-1}$ are in $\langle \tau(f_1), \ldots, \tau(f_m) \rangle$. Doing this for all $d \geq 1$, we get $g' \in \langle \tau(f_1), \ldots, \tau(f_m) \rangle$, hence $g = \tau^{-1}(g') \in \langle f_1, \ldots, f_m \rangle = I$. This finishes the proof. ∎